\documentclass[reprint]{revtex4-1}
\usepackage{amsmath, amsfonts, amssymb, fullpage, graphicx}

\begin{document}
\title{Non--Equilibrium Transport in Open Quantum Systems via Dynamically Constrained non--Hermitian Boundary Domains}
\author{Justin E. Elenewski$^1$, Yanxiang Zhao$^2$, and Hanning Chen$^1$}
\affiliation{$^1$Department of Chemistry and $^2$Department of Mathematics\\ The George Washington University, Washington, DC 20052.}
\email{chenhanning@gwu.edu}

\begin{abstract}
The accurate simulation of real--time quantum transport is notoriously difficult, requiring a consistent scheme to treat incoming and outgoing  fluxes at the boundary of an open  system. We demonstrate a method to converge non--equilibrium steady states using non--Hermitian source and sink potentials alongside the application of dynamic density constraints during wavefunction propagation.  This scheme adds negligible cost to existing computational methods while exhibiting exceptional stability and numerical accuracy. 
\end{abstract}

\maketitle

\section{Introduction}

\par Quantum transport in an open system is characterized by a persistent non--equilibrium state, as maintained through the continuous exchange of particles with a pair of external reservoirs held at distinct chemical potentials \cite{Frensley1990, Ferry2009}.  When other couplings between the reservoirs and the scattering region are negligible, distinct conductance channels form within the quantum device.  This situation is well--described through the Landauer formalism \cite{Landauer1957, Landauer1970, Fisher1981, Buttiker1985, Buttiker1986}, in which each channel contributes a quantum conductance of $\mathcal{G} = (2e^2 / \hbar) T$ to transport, as defined through the transmission coefficient $T$ for the scattering region.  

\par  Practical calculations within the Landauer formalism may be conducted using a variety of methods, including direct determinations of the $S$--matrix via finite difference schemes \cite{Akis2013} and through explicit Monte Carlo simulations of electronic transport.  The latter simulations are readily conjoined with open boundary conditions to eliminate finite size effects in a closed system \cite{Crandall1989, Santos1998, Gonzalez1996} and with Poisson solvers \cite{Laux2004} to explicitly treat the temporal evolution of electrostatics during transport.    This framework must be extended when the quantum subsystem is strongly coupled to the reservoirs or when an explicit treatment of dissipation and time--dependent excitation is required \cite{Oriols2013}.  In these cases, schemes that directly propagate the Schr\"{o}dinger equation in time \cite{Register1991, Arnold2006, Arnold2008, Arnold2008b} or that include these effects directly within quantum master equations \cite{Knezevic2013} may become preferable.

\par Atomic-- or molecular--scale contacts are a particular challenge, as any accurate description requires the use of an electronic structure method such as density functional theory (DFT) \cite{Hohenberg1964, Kohn1965}.  A qualitative model for transport may be obtained by conjoining ground--state DFT with the non--equilibrium Green's function (NEGF) framework \cite{Stefanucci2004, DiVentra2000, Xue2001,  Xue2002}. Unfortunately, this method overestimates the electrical conductance by one to two orders of magnitude, limiting applications for practical electronic devices.  This shortcoming arises from a combination of factors, including the inability of steady--state NEGF+DFT calculations to explicitly treat the time--dependent evolution of Kohn--Sham orbitals during charge migration \cite{Gorling1995, Evers2004, Koentopp2008}.  Alternatively, the electronic wavefunctions may be directly propagated using real--time time--dependent DFT (RT--TDDFT) \cite{Yabana1996} or Ehrenfest dynamics \cite{Li2005}.  While adequately capturing dynamic phenomena, this approach is prone to unphysical particle accumulation and reflection at simulation boundaries \cite{Qian2006, Varga2011}.

\par The undesirable boundary effects associated with a finite system may be partially eliminated through the complex absorbing potential (CAP) method \cite{Neuhasuer1989, Vibok1992, Vibok1992b,Brouard1994,Riss1996,Ge1997,Palao1998,Ferry1999,Manolopoulos2002,Poirier2003,Poirier2003b, Muga2004} in which an imaginary potential term $\hat{V}_C = i\Gamma$, $\Gamma \in \mathbb{R}$ is added to the electronic structure Hamiltonian $\hat{H}_\text{ES}$, with the requirement that this potential is only nonzero at the periphery of the simulation \cite{Ferry1999}.  The new Hamiltonian $\hat{H}' = \hat{H}_\text{ES} + \hat{V}_C$ then affords a propagator with an additional non--unitary modulation $\psi(x,t + \Delta t) = \exp [-i\hat{H}_\text{ES} \Delta t / \hbar] \exp [\Gamma \Delta t / \hbar] \psi(x,t)$ of the wavefunction amplitude during time evolution \cite{Kosloff1986, Neuhasuer1989,Brouard1994, Palao1998, Manolopoulos2002, Muga2004}.   If the strength of the imaginary potential $\Gamma = \Gamma^- < 0$, then the probability density $\rho(x,t) = \psi^*(x,t) \psi(x,t)$ will decrease under the influence of $\Gamma$ as $e^{-2\vert\Gamma\vert t / \hbar}$, emulating loss of particles from the system through a `sink.'  Conversely, if $\Gamma = \Gamma^+ > 0$, the particle density will increase, mimicking an incoming electron flux generated by a boundary `source.'  While $\hat{V}_C = i\Gamma^-$ is sufficient to abrogate artifacts associated with scattering from the simulation boundaries in NEGF calculations \cite{Driscoll2008, Varga2009, Feldman2014},  the uncompensated loss of particles will ultimately retard transport during time--dependent simulations \cite{Varga2011}.  Attempts to avoid this by introducing a compensatory generating potential typically fail or require elaborate algorithmic constructions due to the difficulty in maintaining parity between particle creation and annihilation \cite{Wibking2012}.  As such, the tandem use of source and sink potentials has only been successful in the time--independent case \cite{Varga2007, Wahlstrand2014}.  While efficacious methods have been developed to directly inject a wavefunction during time--dependent propagation, they require modifications to an electronic structure method beyond the introduction of simple potential terms \cite{Wibking2012, Reddy2012}. 

\par To circumvent these limitations, we have developed a simple, self--consistent method to balance the loss and gain profile by applying dynamic density constraints within the source region during real--time propagation.  While static constraints are well established for time--dependent quantum calculations \cite{Umar1985, Wu2005, Oberhofer2009}, we demonstrate that their adaptive refinement during real--time propagation self--consistently converges the system to a fixed non--equilibrium steady state.  This method has broad applicability for the simulation of quantum transport, particularly when using electronic structure methods that admit a single--particle, mean--field description.   

\section{Complex Potentials}

\par   The use of a complex boundary potential may be generalized to a certain limit of multiparticle systems by following the single--particle construction. Let $\hat{V}_C(x)$ be a complex potential that is compactly supported on a subdomain $\mathcal{V}_C \subset \mathbb{R}^3$ and let this potential vanish elsewhere in space.  Furthermore, assume that the Hamiltonian $\hat{H} = \hat{H}_0 + \hat{V}_C$ describes an $N$--particle system that, outside of $\mathcal{V}_C$, is characterized by a conventional Hermitian Hamiltonian 

\begin{equation}
\hat{H}_0 = -\sum_{i=1}^N \frac{\hbar^2 \nabla_i^2}{2 m_i} + \hat{V}_0(x_1, \dots, x_N),
\end{equation}

\noindent with $\hat{V}_0(x_1, \dots, x_N)$ a many--body potential, $m_i$ the mass of the $i$--th particle, and $\nabla_i$ the gradient with respect to its coordinate frame.  We are interested in a situtation analogous to DFT, in which a many--body system is mapped to a set of non--interacting particles evolving in an effective potential 

\begin{equation}
\hat{V}_0(x_1, \dots, x_N) = \sum_{i=1}^N \hat{V}_{\text{eff},i}(x_i),
\end{equation}

\noindent that emulates the multiparticle couplings.  In this `mean--field' situation the total particle number $\mathcal{N}(t)$ is simply a sum of the individual single--particle densities  $\mathcal{N}(t) = \sum_{i=1}^N \int_{\mathbb{R}^3} \rho_i(x,t) \, dV$.  The time--dependent Schr\"{o}dinger equation then gives an equation of motion for each $\rho_i(x,t)$, admitting a physically intuitive evolution for the net particle number  

\begin{equation}
\begin{split}
\partial_t \mathcal{N}(t) & =  \partial_t \sum_{i=1}^N \langle  \psi_i(x,t) \vert \psi_i(x,t) \rangle \\
&= \frac{1}{i\hbar} \sum_{i=1}^N \langle \psi_i(x,t) \vert (\hat{H} - \hat{H}^\dagger)\vert \psi_i(x,t) \rangle \\
&= \frac{2}{\hbar} \sum_{i=1}^N \langle \psi_i(x,t) \vert \text{Im}[\hat{V}_C(x)] \vert \psi_i(x,t)\rangle,
\end{split}
\end{equation}

\noindent where we have adopted notation so that $\hat{H}^\dagger$ is the Hermitian conjugate of $\hat{H}$.   If we assume the simple case of a constant imaginary potential $\hat{V}_C (x) = i\Gamma$, the norm associated with the $i$-th wavefunction $\psi_i(x,t)$ will evolve in a manner determined solely by the portion of its density contained within $\mathcal{V}_C$ so that

\begin{equation} \label{ImDensEvol}
\partial_t \mathcal{N}_i(t) = \partial_t \int_{\mathbb{R}^3} \rho_i(x,t) \, dV = \frac{2 \Gamma}{\hbar}\int_{\mathcal{V}_C} \rho_i(x,t) \, dV,
\end{equation}

\noindent since the potential, and hence the expectation value, vanish outside of this region.  Equating terms under the integrals (\ref{ImDensEvol}) affords the expected time dependence $\rho_i(x,t) = \rho_i(x,0) \exp (2\Gamma t / \hbar)$ for the density. It should be noted that $\rho(x,t)$ must be nonzero on at least one point within $\mathcal{V}_C$ if the evolution defined by $\hat{V}_C(x)$ is to occur.

\par A word of caution is required when using these non--Hermitian effective potentials.  A particle sink term $\Gamma < 0$ is always physically well--defined, even within a multiparticle system, since the norms $\mathcal{N}_i(t)$ monotonically decrease as $t \longrightarrow \infty$ and hence $0 \leq \mathcal{N}_i(t) \leq \mathcal{N}_i(0) = 1$ at all times. Conversely, a source term $\Gamma > 0$ may become pathological as $\mathcal{N}_i(t)$ is unbounded from above and thus permits unphysical state occupancies \cite{Elenewski2014}.  Nonetheless, the point is clear: in an open system the sink potential may be used to mimic an outgoing boundary, while the source potential corresponds to an incoming particle flux.

\par  We require one final result before we may exploit $\hat{V}_C$ algorithmically. In the presence of a non--Hermitian potential $\hat{V}_C(x)$, the continuity equation relating  $\partial_t \rho_i(x,t)$ and the probability current density $\vec{j}_i(x,t) = (\hbar / m) \text{Im} [\psi_i^*(x,t) \nabla\psi_i(x,t)]$ is modified, leading to an integral expression:  

\begin{align} \label{continuity}
\frac{\partial}{\partial t} \int_{\mathcal{V}} \rho_i(x,t) \, dV &+  \int_{\mathcal{V}} \nabla \cdot \vec{j}_i(x,t)\, dV\\ &= \frac{2}{\hbar} \int_{\mathcal{V}} \text{Im}[\hat{V}_C(x)] \rho_i(x,t)\,dV,
\end{align}

\noindent for some arbitrary compact volume $\mathcal{V} \subset \mathbb{R}^3$ \cite{Ferry1999}.  The first term is associated with a growth in, or reduction of, the number of particles contained within $\mathcal{V}$, while the second term corresponds to the change of particle number within $\mathcal{V}$ due to a flux through the boundary $\partial \mathcal{V}$.  The remaining term is dictated by the non--Hermitian behavior of  $\hat{V}_C(x)$, adding or removing density from the system in a manner that does not conserve particle number.

\begin{figure}\label{figure1}
\begin{center}
\includegraphics[scale=1.0]{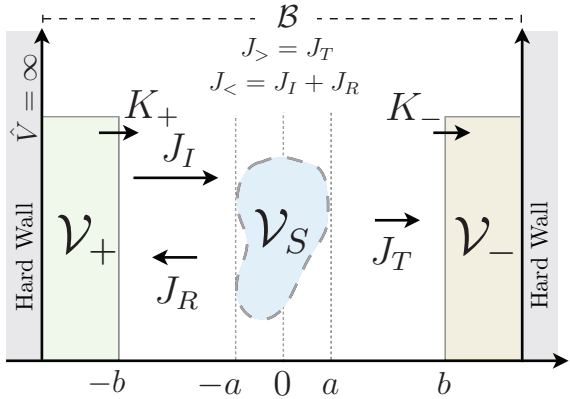} 
\end{center}
\caption{Geometry for transport through a Hermitian scattering region $\mathcal{V}_S$ in the presence of non--Hermitian source and sink potentials. Fluxes are labeled in the regions where they are defined.}
\end{figure}

\section{Open Transport With Balanced Loss and Gain}

\subsubsection{Overall Construction and System Geometry}

\par The mathematical properties of complex boundary potentials allow us to emulate an open quantum system in a manner that is compatible with mean--field electronic structure methods.  In fact, our simplistic constructions, which require only the physics inherent to DFT itself, are designed to be readily integrated into an existing code base.   For pedagogical simplicity we assume a two--terminal device, however, our method is readily extended to multiterminal configurations that contain an arbitrary number of incoming and outgoing particle fluxes.

\par  To begin, we assume that our system of interest occupies a simply connected region $\mathcal{B} \subset \mathbb{R}^3$ and that the scattering process is described by a Hermitian potential contained within a subregion $\mathcal{V}_S \subset \mathcal{B}$ (Fig. 1).  To mimic the incoming and outgoing leads, we construct a pair of nonintersecting regions $\mathcal{V}_+, \mathcal{V}_- \subset \mathcal{B}$ on either side of $\mathcal{V}_S$, corresponding to an asymptotic regions of the terminals where particle injection and removal occur.  As a further requirement, and to ensure this asymptotic separation, we mandate that $\mathcal{V}_{S}$ is separated from $\mathcal{V}_+$ and $\mathcal{V}_-$ by a potential--free region where the scattering states undergo propagation as free particles.  Following this convention we define the source $\hat{V}_+ = i\Gamma^+$ and sink $\hat{V}_- = i\Gamma^-$ potentials within $\mathcal{V}_+$ and $\mathcal{V}_-$.  Finally, we presume that the transport dynamics are generated by  an external bias potential $\hat{V}_\text{ext}$, yielding a net current $\vec{j} = \sum \vec{j}_i (x,t)$ that is directed from $\mathcal{V}_+$ to $\mathcal{V}_-$(Fig. 1).   

\par Due to the external field, a particle flux $J_I$ will impinge on $\mathcal{V}_S$ from the direction of the source lead $\mathcal{V}_+$.  A fraction $T = J_T / J_I $ of this flux  will be transmitted through the scattering potential and subsequently absorbed within $\mathcal{V}_-$, while the remainder $R  = \vert J_R / J_I\vert = 1 - (J_T / J_I)$ is reflected back toward $\mathcal{V}_+$.    If the fluxes $J_T$ and $J_R$ entering $\mathcal{V}_-$ and $\mathcal{V}_+$ at any given instant are immediately reemitted from $\mathcal{V}_+$ as a new flux of magnitude $\vert J_R\vert + \vert J_T\vert$, the density leaving the system will be continuously replenished and particle number will be conserved within the Hermitian region located between the source and drain domains.  As a steady state is approached, the fluxes $J_<$ and $J_>$ on the incoming $(x < -a)$ and outgoing $(x > a)$ sides of the scattering potential will become equal and quantities characterizing a non--equilibrium stationary state, such as the conductance $\mathcal{G} = (e^2 / \pi \hbar) \vert J_T / J_I\vert^2$, are well--defined \cite{Imry1999}. The non--Hermitian boundary potentials allow us to employ a hard wall at $\partial \mathcal{B}$, ensuring compatibility with numerical methods while giving a configuration that resembles an open quantum system.  
 
\par In the steady state, particle density must not accumulate at any point.  To enforce this requirement, the rates of particle addition and removal must be balanced so that $\partial_t \mathcal{N}_- (t) = -\partial_t \mathcal{N}_+ (t)$, where $\mathcal{N}_\pm (t) = \sum_{i=1}^N \int_{\mathcal{V}_\pm} \rho_i(x,t) \, dV$ is the net norm in the source or sink region.  Since we  assume a constant imaginary potential within the non--Hermitian regions, these rates will  depend on the net densities contained within the source and sink domains $\mathcal{V}_+$ and $\mathcal{V}_-$ as well as on the magnitudes of $\Gamma^-$ and $\Gamma^+$ (Eq. \ref{ImDensEvol}).  Due to the inherent difficulty in directly balancing the potential strengths, our strategy is to set $\Gamma = \Gamma^+ = -\Gamma^-$ and instead manipulate the density in the generating region. For numerical purposes, we require that propagation is discretized into timesteps of width $\Delta t$, with the time at step $k$ given by $t_k = k \,\Delta t$.  Furthermore, to ensure that wavefunction norms are well--defined, we assume that the initial wavefunctions are a superposition of Gaussian wavepackets.

\subsubsection{Sink Potential}

\par Assume that a wavepacket $\psi(x,t)$ has been transmitted through the scattering region $\mathcal{V}_S$, resulting in a right--moving  current distribution $\vec{j}_T (x,t) = \vec{j}(x,t)\vert_{x > a}$ in the outgoing terminal.  If we adopt a coordinate system where the preferred axis vector $\hat{n}$ is parallel to the net current,  then the transmitted flux is given by $J_T(x,t) = \int_{D(x)} \vec{j}_T(x',t) \cdot d\vec{A}$, where $D(x)$ is a dividing plane perpendicular to $\hat{n}$ that also contains the point $x$.  For  simplicity we assume that the potential $\hat{V}_-$ is reflectionless, and hence that there is no leftmoving current in this region ($J_>(x,t) = J_T(x,t)$).   The transmitted particle will ultimately pass through the boundary $\partial \mathcal{V}_-$ and enter $\mathcal{V}_-$  at a timestep $t_k$, giving a flux $K_-(t_k) = J_>(b,t_k)$ into the sink region.  If $\vert \Gamma^- \vert$ is sufficiently large, the density introduced into $\mathcal{V}_-$ will rapidly decrease to a negligible value and will not reemerge into the scattering or free propagation domains.

\par To establish a relation between $K_-(x)$ and the rate at which particles are removed from the Hermitian region $\mathcal{V}_H = \mathcal{B} \setminus (\mathcal{V}_- \cup \mathcal{V}_+)$, temporarily assume that no source potential is present ($\Gamma^+ = 0$).  Under these conditions,  probability density in $\mathcal{B}$ may be lost  when it is attenuated by the sink potential $\hat{V}_-$ or when it escapes through  boundary $\partial \mathcal{B}$ surrounding the simulation.  The latter case is impossible due to our hard--wall confining potential, leaving only the sink term to modulate the simulation volume.  Since the potential $\hat{V}_-$  vanishes outside of $\mathcal{V}_-$,  the rate $\partial_t \mathcal{N}_\mathcal{B}(t)$ at which particles are removed from $\mathcal{B}$ must be equal to the rate $\partial_t \mathcal{N}_{\mathcal{V}_-}(t)$ at which they are absorbed within $\mathcal{V}_-$.  Furthermore, density may only leave the Hermitian region $\mathcal{B} \setminus \mathcal{V}_-$ external to $\mathcal{V}_-$  by passing through the joint boundary $\partial \mathcal{V}_-$  into $\mathcal{V}_-$.  If $\Gamma^-$ is assumed to be arbitrarily large, the $\partial \rho / \partial t$ term in the continuity equation (\ref{continuity}) becomes negligible in $\mathcal{V}_-$ due to the rapid removal of particle density.  Under these conditions, no flux reemerges from $\mathcal{V}_-$ and no particles accumulate in this region, implying that the rate of particle removal from $\mathcal{B} \setminus \mathcal{V}_-$ is precisely matched by the rate of absorption in $\mathcal{V}_-$ and hence  $\partial_t \mathcal{N}_{\mathcal{B} \setminus \mathcal{V}_-} = \partial_t \mathcal{N}_{\mathcal{V}_-}$.  Using Stokes' theorem and noting that $\partial_t \mathcal{N}_{\mathcal{B} \setminus \mathcal{V}_-} = J_> (b,t)$ and $J_> (b,t) = \int_{\partial \mathcal{V}_-} \vec{j}_{T}(x,t) \cdot dA$, we have that 

\begin{eqnarray}
K_-(t) &=& \int_{\partial \mathcal{V}_-} \vec{j}_T(x,t_k) \cdot d\vec{A} \\
&=& -\frac{2 \Gamma^-}{\hbar} \int_{\mathcal{V}_-} \rho(x,t_k) \, dV
\end{eqnarray}

\noindent meaning that the flux through the outgoing boundary is  determined entirely by the magnitude of $\Gamma^-$ and the density contained within $\mathcal{V}_-$. Thus, within this construction, $\mathcal{V}_-$ behaves as an open, outgoing boundary for the Hermitian region.

\subsubsection{Source Potential}

\par To maintain a steady current, we must now construct a compensatory flux $K_+(t_{k+1}) = K_-(t_k)$ that is emitted from $\mathcal{V}_+$ at a subsequent timestep.  Since we set $\Gamma_+ = -\Gamma_-$, this may be implemented by constraining the density within $\mathcal{V}_+$ so that

\begin{equation}
 \int_{\mathcal{V}_+} \rho(x,t_{k+1}) \, dV =  \int_{\mathcal{V}_-} \rho(x,t_k) \, dV.
\end{equation}

\noindent which implies that $\partial_t \mathcal{N}_+ = -\partial_t \mathcal{N}_-$, and hence the net particle number must remain constant.   In practice, we must also account for the flux $J_R(x,t) = \int_{D(x)} \vec{j}_R(x,t) \cdot d\vec{A}$  that is reflected back from the scattering region $\mathcal{V}_S$ toward $\mathcal{V}_+$ and ensure that this density is reintroduced into the system.  This construction assumes that the reflected flux $K_R(t_k) = J_R(-b,t_k)$ may be efficiently estimated using some numerical scheme.  The constraint within $\mathcal{V}_+$ is then revised so that the target density affords an outgoing flux

\begin{eqnarray}
K_+(t_{k+1}) &=& K_-(t_k) - K_R(t_k) \\
&=& -\frac{2 \Gamma^+}{\hbar} \int_{\mathcal{V}_+} \rho(x,t_k) \, dV
\end{eqnarray}

\noindent that keeps the net particle number in $\mathcal{V}_H$ constant.  Since the rate of particle generation is proportional to the net density within $\mathcal{V}_+$, care must be taken to prevent the source region from `blowing up.'  In this case, we require that the surface--to--volume ratio of $\mathcal{V}_+$ is sufficiently large, so that any density generated in $\mathcal{V}_+$ rapidly emerges as an outgoing flux before it may accumulate ($\partial \rho / \partial t \approx 0$). For a wavepacket with group velocity $v_g = \hbar q / m$, this implies that the generating region must have a characteristic length scale on the order of $\ell \sim v_g \, \Delta t$ along the direction of current flow.  This is easily estimated by assuming that $q \sim k_F$, where $k_F$ is the Fermi wavevector for the system of interest, since most current carrying states will lie in the vicinity of the Fermi level.

\par One final provision is related to the geometry of $\mathcal{V}_+$ itself.  If the domain is slab--like, any flux generated in  $\mathcal{V}_+$ may depart through either the surface facing $\mathcal{V}_S$ or the surface facing away from the scattering region.  This situation is easily resolved by placing an infinite potential along the outer boundary of $\mathcal{V}_+$ or by requiring that the pathological boundary of $\mathcal{V}_+$ coincide with $\partial\mathcal{B}$.   This need is further mitigated by the presence of an external bias potential, which encourages migration of density toward the scattering domain. 

\begin{figure}\label{figure2}
\begin{center}
\includegraphics[scale=1.0]{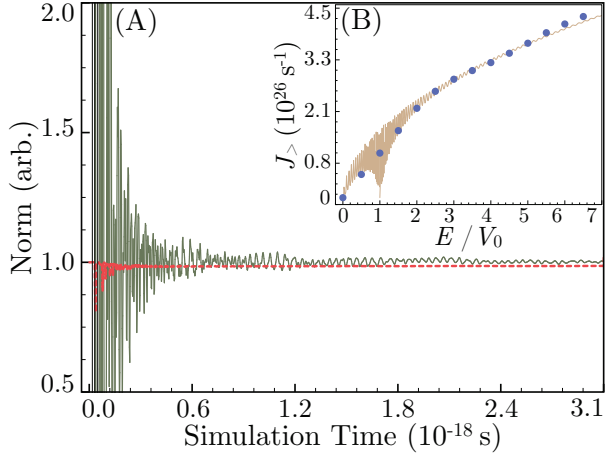} 
\end{center}
\caption{Convergence of the norm in the Hermitian region (dashed pink line) and the flux ratio $\vert J_> \vert /\vert J_<\vert$ (solid green line) during real--time propagation of an initial wavepacket state.  Inset: Scaling of the outgoing flux $J_>(E)$ as a function of the incident packet energy (blue circles) alongside the rescaled analytical current for plane wave scattering off a square barrier (beige line).  Fluxes are averaged over a single period at $2.76 \times 10^6$ timesteps ($1.7 \times 10^{-18}$ s). }
\end{figure}

\section{Numerical Methods}

\par To assess the computational validity of our approach, we employ a one--dimensional, single--particle system, with simulation parameters chosen to accelerate calculations for demonstrative purposes.  Natural atomic units are assumed for all numerical work in this manuscript ($\hbar = m_e = e = 1$, where $m_e$ is the electron mass and $e$ is the fundamental unit of electronic charge).  The wavefunctions are propagated in real--time using the forward Euler method for a complex potential, as described elsewhere, with timestep of $\Delta t = 2.5 \times 10^{-8}$ a.u. $ (6.1 \times 10^{-25}$ s) and spatial discretization of $\Delta x = 5 \times 10^{-4} \, a_0$  (26.5 fm) for a total of $5 \times 10^6$ steps ($3.1 \times 10^{-18}$ s) \cite{Elenewski2014}.  Propagation occurs on the interval $x \in [0.0, 2.0\, a_0]$, measured in units of the Bohr radius $a_0$, with a hard boundary that is flanked by constant source and sink potentials ($\mathcal{V}_- = (0.0,0.05\, a_0]$; $\mathcal{V}_+ = [1.9\, a_0,2.0\, a_0)$; $\Gamma^+ = -\Gamma^- =   2.5 \times 10^4$ Ha $ ( 0.68$ MeV)).  The potential strength is chosen so that any density entering  $\mathcal{V}_-$ will be rapidly absorbed and thus ensuring that no particles will be reflected into the Hermitian region.  The scattering center is treated as a rectangular barrier of height $V_S = V_0 = 1.0 \times 10^5$ Ha (2.76 MeV) defined for $x_0 \in [0.9\, a_0, 1.1\, a_0]$, and thus corresponding to the center of the scattering region $\mathcal{V}_S = (0.05\, a_0, 1.9\, a_0)$.

\subsubsection{Density Constraints and Flux Determination}

\par During real--time propagation, the density in the source region $\mathcal{V}_+$ is constrained by setting 

\begin{multline}\label{seedpacket}
\psi(x,t_k)\vert_{x \in \mathcal{V}_+} =\\ [\mathcal{N}_- (t_{k-1}) + \mathcal{N}_R (t_{k-1})]^{1/2}\, \psi_G(x,t) 
\end{multline}

\noindent after every $N_R = 800$ timesteps (every $4.8 \times 10^{-22}$ s), where $\mathcal{N}_- (t_{k-1})$ is the total density absorbed within $\mathcal{V}_-$ at the preceding timestep, $\mathcal{N}_R (t_{k-1})$ is an estimator for the density transferred into $\mathcal{V}_{+}$ due to left--moving reflected current, and  $\psi_G(x,t)$ is a normalized Gaussian packet 

\begin{equation} \label{gausspacket}
\psi_G(x,0) =  (\pi \sigma^2)^{-1/4} e^{i q_0 (x-x_0)} e^{-(x-x_0)^2 / 2\sigma^2}
\end{equation}

\noindent that serves as a `seed' wavefunction.  The choice of this function is arbitrary, amounting to a local choice of basis for the system.  The seed is centered at the midpoint of $\mathcal{V}_+$ and assigned an initial a width $\sigma^2$ to ensure that $\psi_G(x,0)$ falls below machine precision and hence is effectively zero at the boundary ($x_0 = 0.025\, a_0$, $q_0 = 1000 \, a_0^{-1}$, $\sigma^2 = 1 \times 10^{-4} \, a_0^2$).  The  wavevector $q_0$ is set equal to that of the initial wavefunction in the scattering region, emulating the effect of an external field.  For the simple case herein, the scaling prefactor for the seed wavepacket (\ref{seedpacket}) is equal to the norm removed from the Hermitian scattering region 
  
\begin{equation}
\mathcal{N}_- (t_{k-1}) + \mathcal{N}_R (t_{k-1}) = 1 - \mathcal{N}_H (t_{k-1})
\end{equation}
  
\noindent where $\mathcal{N}_H(t)$ is the norm within the Hermitian subsystem $\mathcal{V}_H = \mathcal{B} \setminus (\mathcal{V}_+ \cup \mathcal{V}_-)$.  The fluxes on the incoming $J_>$ and outgoing $J_<$ sides of the scattering center are determined by calculating the mean flux within windows of 0.15 $a_0$ units in width centered about the midpoint of the upper and lower free propagation regions, respectively.  Note that the delay period $N_R$ is necessary to ensure that the generated component of the wavefunction will acquire a nonzero expectation value for its momentum and hence continue to propagate within the scattering region.

\subsubsection{Gaussian Wavepacket}

\par As a worst--case benchmark, we consider a scenario in which the initial scattering wavefunction is a tightly--localized Gaussian wavepacket.  This corresponds to the largest possible deviation from a steady--state density distribution, accompanied by a broad wavevector dispersion during propagation.  In this case we set $\psi(x,0) =  (\pi \sigma^2)^{-1/4} e^{i q_0 (x-x_0)} e^{-(x-x_0)^2 / 2\sigma^2}$ and  center the wavefunction at $x_0 = 0.5 \, a_0$ with wavevector $q_0 = 1000 \, a_0^{-1}$ and width $\sigma^2 = 5.0 \times 10^{-3} \, a_0^2$.  The choice of a large wavevector was chosen to accelerate our benchmark calculations, however, the calculation may be repeated for any other $q_0$, requiring at most a slight adjustment of update parameters for the generating potential.  Even lacking such optimization, the method is highly transferrable to a broad range of initial wavevectors, and hence external biases, as discussed below. 

 \par During these simulations, the ratio of transmitted flux to incident flux $ \vert J_> \vert / \vert J_< \vert$ (mean value = 0.987, standard deviation = 0.012)  and the total norm in the Hermitian scattering region $\mathcal{N}_H$ (mean value = 0.989, standard deviation = 0.001) become quasi--stationary after $2 \times 10^6$ steps ($1.22 \times 10^{-18}$ s), representative of a worst--case scenario for convergence due to the tightly localized initial density (Fig. 2). The periodic modulation observed in these quantities, as well as their departure from unity, is a consequence of the constraint update procedure and the inherent asymmetry in  arrival times at the boundary for transmitted and backscattered densities.  As expected, the packet momentum distributions broaden during propagation, concomitant with wavepacket spread, resulting in bound states trapped between $\mathcal{V}_+$ and $\mathcal{V}_S$ with $E(q) = q^2 / 2 < V_0$.  After $5 \times 10^6$ timesteps ($1.2 \times 10^{-18}$ s), these states contain $\sim 0.16 \, \mathcal{N}_H $ units of bound norm. While present in this toy model, such trapping is not a concern for realistic systems as the bound states would be occupied by non--current--carrying particles.  As a final note, these behaviors are not confounded by backscattering from the outgoing boundary, as our chosen magnitude for the sink potential strength $\Gamma^-$ ensures that nearly all of the outgoing density is absorbed.  Should reflections become problematic, reflection--minimizing absorbing potentials may be adopted \cite{Manolopoulos2002, Muga2004}.

 \par These calculations may be repeated for different initial center--of--packet energies $E_0 = q_0^2 / 2$, and hence different initial wavevectors $q_0$, affording a profile for $J_>(E)$ that scales in parallel to the flux for a plane wave incident on the same barrier $J_{\text{pw}}(E_0) = \rho_0 \sqrt{2E_0}\, T(E_0)$  as calculated using the exact transmission coefficient $T(E_0)$.  A linear least squares minimization of $\vert \alpha J_{\text{pw}}(E_0) - J_>(E_0)\vert$ gives a scaling coefficent  of $\alpha = 0.225$.  This is comparable to the estimate of $\alpha_\text{est.} = 0.353$ given by assuming a factor of $0.84$ for both freely propagating and energetically competent states in our simulation, alongside an even distribution of one unit of norm throughout the system.  An exact correspondence is not expected, since the system contains a superposition of plane--wave states due to the increasingly broad wavevector dispersion, however, the scaling behavior as a function of the initial $q_0$ should parallel the plane wave--case.  Furthermore, since convergence is limited by the spread of the wavepacket throughout the simulation cell, requiring a homogenous distribution for $\vert J_> / J_< \vert \sim 1$, the choice of a different wavevector has a trivial effect on the rate of convergence.  As cross--validation, we have obtained similar results using a fourth--order Runge--Kutta propagator and a plane--wave functional form for the source (not shown).
 
 \begin{figure}\label{figure3}
\begin{center}
\includegraphics[scale=1.0]{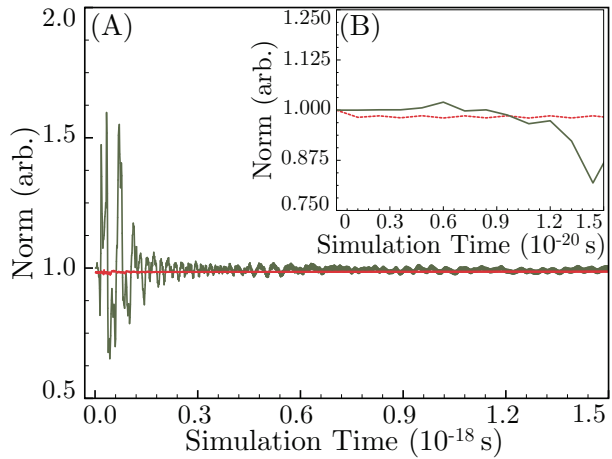} 
\end{center}
\caption{Convergence of the norm in the Hermitian region (dashed pink line) and the flux ratio $\vert J_> \vert /\vert J_<\vert$ (solid green line) starting from an extended plane--wave state.  Inset: Convergence of the aforementioned quantities for short timescales.  }
\end{figure}

 \subsubsection{Plane wave states}
 
 \par In a more realistic scenario, the initial wavefunction will be an eigenstate of the electronic structure Hamiltonian.  As such, we performed an additional set of calculations by adopting an exact eigenstate for plane wave scattering at a rectangular potential barrier as our initial condition:  
 
\begin{equation}
\psi(x, t = 0) =  \begin{cases}
e^{i q_0 x} + B e^{-i q_0 x} & x \leq -b \\
C e^{iq_0 x} + D e^{-i q_0 x} & -b \leq x \leq b \\
 F e^{iq_0 x} & x \geq b,
\end{cases}
\end{equation}
 
 \noindent where the coefficients are determined by enforcing continuity of $\psi(x,0)$ and $\partial_x \psi(x,0)$ between each region space ($q_0 = 1000 \, a_0^{-1}$). Convergence is obtained within $5 \times 10^5$ timesteps ($3.1 \times 10^{-19}$ s), leading to a  persistent non--equilibrium steady--state (Fig 3A; Table I).  Even before this point, the maximal fluctuation in  $\vert J_> / J_< \vert$ is substantially smaller than the wavepacket simulations, with the flux ratio bounded by $0.6 \leq \vert J_> / J_< \vert \leq 1.6$ at all times.  The large oscillations observed before $t = 5 \times 10^5$ timesteps ($3.1 \times 10^{-19}$ s) are a transient consequence of multiple reflections from the rectangular barrier, and the small scale oscillations observed thereafter are related to the update frequency and generating packet shape.  When examining the short time limit (Fig. 3B), the system exhibits a particularly high degree of stability up to $2.0 \times 10^4$ timesteps ($1.20 \times 10^{-20}$ s; mean value = 0.983, standard deviation = 0.003), corresponding to an interval with sufficient sampling for current determinations.   In fact, below $1.0 \times 10^4$ timesteps ($6.1 \times 10^{-21}$ s) the norm and ratio of $\vert J_> / J_< \vert$  are essentially constant, ensuring that the system would be stable for any realistic, DFT--based simulations.  
 
 \par It should be noted that, even for  long--timescale  non--equilibrium steady--states ($k \geq 5 \times 10^5$ timesteps; $3.1 \times 10^{-19}$ s), the magnitudes of charge and current fluctuations observed using our method are comparable to the charge depletion in a normal RT--TDDFT  transport simulation \cite{Varga2011} with a periodicity that is a comparable fraction of total simulation time.  As such, the update protocol should not affect any physical behavior,  even if conjoined with an electronic structure method, as oscillations of this magnitude do not alter the Hartree term strongly enough to have a quantitatively meaningful effect.   We reemphasize that the parameters we have adopted for our reference calculations are intended to be illustrative and provide a rapid demonstration of our method.  Nonetheless, the extension to practical RT--TDDFT calculations and higher--order propagator schemes is feasible.
 
 \begin{table}
 \begin{tabular}{l||c|c|c|c}
 Time ($10^{-18}$ s) & $\overline{\mathcal{N}_H}$ & $\sigma(\mathcal{N}_H)$ &  $\overline{\vert J_> / J_< \vert}$ & $\sigma(\vert J_> / J_< \vert)$ \\
 \hline\hline
$ t_k \leq 0.0061 $ & 0.983 & 0.003 & 1.005 & 0.008 \\
$ t_k \leq 0.012$ & 0.983 & 0.003 & 0.995 & 0.015 \\
$ 0.30 \leq t_k \leq 1.5$ & 0.986 & 0.001 & 0.993 & 0.011 \\
$ 0.060 \leq t_k \leq 1.5$ & 0.986 & 0.001 & 0.993 & 0.010 \\
  \end{tabular}
  \caption{Convergence of parameters associated with propagating an initial plane--wave state.  Values correspond to the mean $\overline{\mathcal{N}_H}$ and standard deviation $\sigma(\mathcal{N}_H)$ of the norm $\mathcal{N}_H$ in the Hermitian region, averaged up to or between the indicated intervals.  Also enumerated are the same parameters for the ratio $\vert J_> / J_< \vert$, reflecting the balance of incoming and outgoing fluxes.}
 \end{table}

 \section{Conclusion and Extensions}

\par Through the judicious use of complex source and sink potentials, we have devised a scheme that can converge non--equilibrium steady states during real time propagation.   Our method is conceptually similar to the Landauer approach and, in some sense, represents a practicable and self--consistent adaptation to coherent, time--dependent electronic structure calculations.    While our numerical tests were limited to the two--terminal case, these principles may be extended to an arbitrary multiterminal device given a suitable estimator for the reflected flux $\mathcal{N}_R (t_{k-1})$.  As such, our results represent an extension beyond  conventional Monte Carlo simulations and other direct optimization methods such as NEGF calculations \cite{Stefanucci2004, DiVentra2000, Xue2001,  Xue2002}, particularly by accommodating transient perturbations such as photoexcitation or  a time--dependent bias potential.  Furthermore, our construction is designed using only  the basic physics inherent to DFT.  It is thus conceptually compatible with the  RT--TDDFT codes \cite{Marques2003, Castro2004, Castro2006, Andrade2012} required for quantitative applications to complex physical systems.  The use of an absorbing potential alone has been practically demonstrated to lengthen the timescale of RT--TDDFT transport simulations \cite{Varga2011}, while manually tuned source and sink potentials can extend static DFT calculations to the open limit \cite{Varga2007}.  Viewed in this context, our construction affords an automated, self--consistent procedure to balance loss and gain in similar procedures, thus providing a tractable scheme for particle injection in DFT calculations \cite{Reddy2012, Wibking2012}.

\par We expect the performance of our algorithm to be enhanced under the  conditions associated with realistic electronic structure calculations.  In the case of higher--dimensional mean--field theories, including DFT,  the effective potential and additional degrees of freedom are expected to mitigate the formation of bound states and ensure homogeneous distribution of density from $\mathcal{V}_+$.  While the application of our method to an orbital--centered basis would be substantially more complicated than the scheme presented herein, our methods apply directly to any real--space, grid--based basis schemes.   Furthermore, since we are only constraining the net density $\mathcal{N}_+ (t) = \int_{\mathcal{V}_+} \Psi(x,t)^* \Psi(x,t)\, dV $, a logical extension to the noninteracting many--particle limit entails copying the wavefunction from $\mathcal{V}_-$ to $\mathcal{V}_+$ during updates, matching the function between regions to maintain boundary continuity, and rescaling its norm to a target value.  Through a systematic use of this procedure, it should be possible to avoid the unphysical state occupancy typically associated with imaginary source potentials \cite{Elenewski2014}.

\par Taken together, these modifications constitute an actionable protocol for quantum transport in open systems at minimal computational expense,  compatible with both contemporary electronic structure methods and mean--field representations of the many--particle problem.  We expect that, with sufficient development, an existing real--time propagation code could be modified by adding imaginary potentials at the incoming and outgoing terminal boundaries of a device. Transport would be initiated by applying an external field, driving electron density into the sink at the outgoing terminal.   By introducing an update procedure for the wavefunction in the source region, and thus periodically enforcing the density constraint, a compensatory flux would ensure steady currents  and conservation of particle number within the simulation cell.  Our technique is expected to be stable on the timescale accessible to RT--TDDFT simulations, making it ideally poised to tackle realistic systems.

\par This research was partially supported by a start--up grant from The George Washington University (GWU). The authors Y. Zhao and H. Chen also appreciate support from the Dean's Interdisciplinary Collaboration Excellence award at GWU.  Computational resources were provided by the Argonne Leadership Computing Facility (ALCF) through the ASCR Leadership Computing Challenge (ALCC) award under Department of Energy Contract DE--AC02--06CH11357 and by the Extreme Science and Engineering Discovery Environment (XSEDE) under National Science Foundation contract TG--CHE130008.


\begin{thebibliography}{60}%
\makeatletter
\providecommand \@ifxundefined [1]{%
 \@ifx{#1\undefined}
}%
\providecommand \@ifnum [1]{%
 \ifnum #1\expandafter \@firstoftwo
 \else \expandafter \@secondoftwo
 \fi
}%
\providecommand \@ifx [1]{%
 \ifx #1\expandafter \@firstoftwo
 \else \expandafter \@secondoftwo
 \fi
}%
\providecommand \natexlab [1]{#1}%
\providecommand \enquote  [1]{``#1''}%
\providecommand \bibnamefont  [1]{#1}%
\providecommand \bibfnamefont [1]{#1}%
\providecommand \citenamefont [1]{#1}%
\providecommand \href@noop [0]{\@secondoftwo}%
\providecommand \href [0]{\begingroup \@sanitize@url \@href}%
\providecommand \@href[1]{\@@startlink{#1}\@@href}%
\providecommand \@@href[1]{\endgroup#1\@@endlink}%
\providecommand \@sanitize@url [0]{\catcode `\\12\catcode `\$12\catcode
  `\&12\catcode `\#12\catcode `\^12\catcode `\_12\catcode `\%12\relax}%
\providecommand \@@startlink[1]{}%
\providecommand \@@endlink[0]{}%
\providecommand \url  [0]{\begingroup\@sanitize@url \@url }%
\providecommand \@url [1]{\endgroup\@href {#1}{\urlprefix }}%
\providecommand \urlprefix  [0]{URL }%
\providecommand \Eprint [0]{\href }%
\providecommand \doibase [0]{http://dx.doi.org/}%
\providecommand \selectlanguage [0]{\@gobble}%
\providecommand \bibinfo  [0]{\@secondoftwo}%
\providecommand \bibfield  [0]{\@secondoftwo}%
\providecommand \translation [1]{[#1]}%
\providecommand \BibitemOpen [0]{}%
\providecommand \bibitemStop [0]{}%
\providecommand \bibitemNoStop [0]{.\EOS\space}%
\providecommand \EOS [0]{\spacefactor3000\relax}%
\providecommand \BibitemShut  [1]{\csname bibitem#1\endcsname}%
\let\auto@bib@innerbib\@empty
\bibitem [{\citenamefont {Frensley}(1990)}]{Frensley1990}%
  \BibitemOpen
  \bibfield  {author} {\bibinfo {author} {\bibfnamefont {W.~R.}\ \bibnamefont
  {Frensley}},\ }\href@noop {} {\bibfield  {journal} {\bibinfo  {journal} {Rev.
  Mod. Phys.}\ }\textbf {\bibinfo {volume} {62}},\ \bibinfo {pages} {745}
  (\bibinfo {year} {1990})}\BibitemShut {NoStop}%
\bibitem [{\citenamefont {Ferry}\ \emph {et~al.}(2009)\citenamefont {Ferry},
  \citenamefont {Goodnick},\ and\ \citenamefont {Bird}}]{Ferry2009}%
  \BibitemOpen
  \bibfield  {author} {\bibinfo {author} {\bibfnamefont {D.~K.}\ \bibnamefont
  {Ferry}}, \bibinfo {author} {\bibfnamefont {S.~M.}\ \bibnamefont {Goodnick}},
  \ and\ \bibinfo {author} {\bibfnamefont {J.}~\bibnamefont {Bird}},\
  }\href@noop {} {\emph {\bibinfo {title} {Transport in Nanostructures (Second
  Edition)}}}\ (\bibinfo  {publisher} {Cambridge University Press},\ \bibinfo
  {year} {2009})\BibitemShut {NoStop}%
\bibitem [{\citenamefont {Landauer}(1957)}]{Landauer1957}%
  \BibitemOpen
  \bibfield  {author} {\bibinfo {author} {\bibfnamefont {R.}~\bibnamefont
  {Landauer}},\ }\href@noop {} {\bibfield  {journal} {\bibinfo  {journal} {IBM
  J. Res. Develop.}\ }\textbf {\bibinfo {volume} {1}},\ \bibinfo {pages} {223}
  (\bibinfo {year} {1957})}\BibitemShut {NoStop}%
\bibitem [{\citenamefont {Landauer}(1970)}]{Landauer1970}%
  \BibitemOpen
  \bibfield  {author} {\bibinfo {author} {\bibfnamefont {R.}~\bibnamefont
  {Landauer}},\ }\href@noop {} {\bibfield  {journal} {\bibinfo  {journal}
  {Phil. Mag.}\ }\textbf {\bibinfo {volume} {21}},\ \bibinfo {pages} {863}
  (\bibinfo {year} {1970})}\BibitemShut {NoStop}%
\bibitem [{\citenamefont {Fisher}\ and\ \citenamefont
  {Lee}(1981)}]{Fisher1981}%
  \BibitemOpen
  \bibfield  {author} {\bibinfo {author} {\bibfnamefont {D.~S.}\ \bibnamefont
  {Fisher}}\ and\ \bibinfo {author} {\bibfnamefont {P.~A.}\ \bibnamefont
  {Lee}},\ }\href@noop {} {\bibfield  {journal} {\bibinfo  {journal} {Phys.
  Rev. B}\ }\textbf {\bibinfo {volume} {23}},\ \bibinfo {pages} {6851}
  (\bibinfo {year} {1981})}\BibitemShut {NoStop}%
\bibitem [{\citenamefont {B\"{u}ttiker}\ \emph {et~al.}(1985)\citenamefont
  {B\"{u}ttiker}, \citenamefont {Imry}, \citenamefont {Landauer},\ and\
  \citenamefont {Pinhas}}]{Buttiker1985}%
  \BibitemOpen
  \bibfield  {author} {\bibinfo {author} {\bibfnamefont {M.}~\bibnamefont
  {B\"{u}ttiker}}, \bibinfo {author} {\bibfnamefont {Y.}~\bibnamefont {Imry}},
  \bibinfo {author} {\bibfnamefont {R.}~\bibnamefont {Landauer}}, \ and\
  \bibinfo {author} {\bibfnamefont {S.}~\bibnamefont {Pinhas}},\ }\href@noop {}
  {\bibfield  {journal} {\bibinfo  {journal} {Phys. Rev. B}\ }\textbf {\bibinfo
  {volume} {31}},\ \bibinfo {pages} {6207} (\bibinfo {year}
  {1985})}\BibitemShut {NoStop}%
\bibitem [{\citenamefont {B\"{u}ttiker}(1986)}]{Buttiker1986}%
  \BibitemOpen
  \bibfield  {author} {\bibinfo {author} {\bibfnamefont {M.}~\bibnamefont
  {B\"{u}ttiker}},\ }\href@noop {} {\bibfield  {journal} {\bibinfo  {journal}
  {Phys. Rev. Lett.}\ }\textbf {\bibinfo {volume} {57}},\ \bibinfo {pages}
  {1761} (\bibinfo {year} {1986})}\BibitemShut {NoStop}%
\bibitem [{\citenamefont {Akis}\ and\ \citenamefont {Ferry}(2013)}]{Akis2013}%
  \BibitemOpen
  \bibfield  {author} {\bibinfo {author} {\bibfnamefont {R.}~\bibnamefont
  {Akis}}\ and\ \bibinfo {author} {\bibfnamefont {D.~K.}\ \bibnamefont
  {Ferry}},\ }\href@noop {} {\bibfield  {journal} {\bibinfo  {journal} {J.
  Comput. Electron}\ }\textbf {\bibinfo {volume} {12}},\ \bibinfo {pages} {356}
  (\bibinfo {year} {2013})}\BibitemShut {NoStop}%
\bibitem [{\citenamefont {Crandle}\ \emph {et~al.}(1989)\citenamefont
  {Crandle}, \citenamefont {East},\ and\ \citenamefont
  {Blakey}}]{Crandall1989}%
  \BibitemOpen
  \bibfield  {author} {\bibinfo {author} {\bibfnamefont {T.~L.}\ \bibnamefont
  {Crandle}}, \bibinfo {author} {\bibfnamefont {J.~R.}\ \bibnamefont {East}}, \
  and\ \bibinfo {author} {\bibfnamefont {P.~A.}\ \bibnamefont {Blakey}},\
  }\href@noop {} {\bibfield  {journal} {\bibinfo  {journal} {IEEE Trans.
  Electron Dev.}\ }\textbf {\bibinfo {volume} {36}},\ \bibinfo {pages} {300}
  (\bibinfo {year} {1989})}\BibitemShut {NoStop}%
\bibitem [{\citenamefont {Santos}\ and\ \citenamefont {Paisana}(3
  44)}]{Santos1998}%
  \BibitemOpen
  \bibfield  {author} {\bibinfo {author} {\bibfnamefont {H.~A.}\ \bibnamefont
  {Santos}}\ and\ \bibinfo {author} {\bibfnamefont {J.~A.}\ \bibnamefont
  {Paisana}},\ }\href@noop {} {\bibfield  {journal} {\bibinfo  {journal}
  {Microlectron. Engr.}\ ,\ \bibinfo {pages} {507}} (\bibinfo {year}
  {43-44})}\BibitemShut {NoStop}%
\bibitem [{\citenamefont {Gonzalez}\ and\ \citenamefont
  {Pardo}(1996)}]{Gonzalez1996}%
  \BibitemOpen
  \bibfield  {author} {\bibinfo {author} {\bibfnamefont {T.}~\bibnamefont
  {Gonzalez}}\ and\ \bibinfo {author} {\bibfnamefont {D.}~\bibnamefont
  {Pardo}},\ }\href@noop {} {\bibfield  {journal} {\bibinfo  {journal}
  {Sol.--State Electron.}\ }\textbf {\bibinfo {volume} {39}},\ \bibinfo {pages}
  {555} (\bibinfo {year} {1996})}\BibitemShut {NoStop}%
\bibitem [{\citenamefont {Laux}\ \emph {et~al.}(2004)\citenamefont {Laux},
  \citenamefont {Kumar},\ and\ \citenamefont {Fischetti}}]{Laux2004}%
  \BibitemOpen
  \bibfield  {author} {\bibinfo {author} {\bibfnamefont {S.~E.}\ \bibnamefont
  {Laux}}, \bibinfo {author} {\bibfnamefont {A.}~\bibnamefont {Kumar}}, \ and\
  \bibinfo {author} {\bibfnamefont {M.~V.}\ \bibnamefont {Fischetti}},\
  }\href@noop {} {\bibfield  {journal} {\bibinfo  {journal} {J. Appl. Phys.}\
  }\textbf {\bibinfo {volume} {95}},\ \bibinfo {pages} {5545} (\bibinfo {year}
  {2004})}\BibitemShut {NoStop}%
\bibitem [{\citenamefont {Oriols}\ and\ \citenamefont
  {Ferry}(2013)}]{Oriols2013}%
  \BibitemOpen
  \bibfield  {author} {\bibinfo {author} {\bibfnamefont {X.}~\bibnamefont
  {Oriols}}\ and\ \bibinfo {author} {\bibfnamefont {D.~K.}\ \bibnamefont
  {Ferry}},\ }\href@noop {} {\bibfield  {journal} {\bibinfo  {journal} {J.
  Comput. Electron}\ }\textbf {\bibinfo {volume} {12}},\ \bibinfo {pages} {317}
  (\bibinfo {year} {2013})}\BibitemShut {NoStop}%
\bibitem [{\citenamefont {Register}\ \emph {et~al.}(1991)\citenamefont
  {Register}, \citenamefont {Ravaioli},\ and\ \citenamefont
  {Hess}}]{Register1991}%
  \BibitemOpen
  \bibfield  {author} {\bibinfo {author} {\bibfnamefont {L.~F.}\ \bibnamefont
  {Register}}, \bibinfo {author} {\bibfnamefont {U.}~\bibnamefont {Ravaioli}},
  \ and\ \bibinfo {author} {\bibfnamefont {K.}~\bibnamefont {Hess}},\
  }\href@noop {} {\bibfield  {journal} {\bibinfo  {journal} {J. Appl. Phys.}\
  }\textbf {\bibinfo {volume} {69}},\ \bibinfo {pages} {7153} (\bibinfo {year}
  {1991})}\BibitemShut {NoStop}%
\bibitem [{\citenamefont {Arnold}\ and\ \citenamefont
  {J\"{u}ngel}(2006)}]{Arnold2006}%
  \BibitemOpen
  \bibfield  {author} {\bibinfo {author} {\bibfnamefont {A.}~\bibnamefont
  {Arnold}}\ and\ \bibinfo {author} {\bibfnamefont {A.}~\bibnamefont
  {J\"{u}ngel}},\ }\enquote {\bibinfo {title} {Analysis, modeling, and
  simulation of multiscale problems},}\ \ (\bibinfo  {publisher} {Springer,
  Berlin--Heidelberg},\ \bibinfo {year} {2006})\ Chap.\ \bibinfo {chapter}
  {Multi--scale modleing of quantum semiconductor devices}, pp.\ \bibinfo
  {pages} {331--363}\BibitemShut {NoStop}%
\bibitem [{\citenamefont {Arnold}\ and\ \citenamefont
  {Schulte}(2008)}]{Arnold2008}%
  \BibitemOpen
  \bibfield  {author} {\bibinfo {author} {\bibfnamefont {A.}~\bibnamefont
  {Arnold}}\ and\ \bibinfo {author} {\bibfnamefont {M.}~\bibnamefont
  {Schulte}},\ }\href@noop {} {\bibfield  {journal} {\bibinfo  {journal} {Math.
  Comput. Sim.}\ }\textbf {\bibinfo {volume} {79}},\ \bibinfo {pages} {898}
  (\bibinfo {year} {2008})}\BibitemShut {NoStop}%
\bibitem [{\citenamefont {Arnold}(2008)}]{Arnold2008b}%
  \BibitemOpen
  \bibfield  {author} {\bibinfo {author} {\bibfnamefont {A.}~\bibnamefont
  {Arnold}},\ }\enquote {\bibinfo {title} {Quantum transport: Modeling,
  analysis, and asymptotics},}\ \ (\bibinfo  {publisher} {Springer,
  Berlin--Heidelberg},\ \bibinfo {year} {2008})\ Chap.\ \bibinfo {chapter}
  {Mathematical Properties of Quantum Evolution Equations}\BibitemShut
  {NoStop}%
\bibitem [{\citenamefont {Knezevic}\ and\ \citenamefont
  {Novakovic}(2013)}]{Knezevic2013}%
  \BibitemOpen
  \bibfield  {author} {\bibinfo {author} {\bibfnamefont {I.}~\bibnamefont
  {Knezevic}}\ and\ \bibinfo {author} {\bibfnamefont {B.}~\bibnamefont
  {Novakovic}},\ }\href@noop {} {\bibfield  {journal} {\bibinfo  {journal} {J.
  Comput. Electron.}\ }\textbf {\bibinfo {volume} {12}},\ \bibinfo {pages}
  {363} (\bibinfo {year} {2013})}\BibitemShut {NoStop}%
\bibitem [{\citenamefont {Hohenberg}\ and\ \citenamefont
  {Kohn}(1964)}]{Hohenberg1964}%
  \BibitemOpen
  \bibfield  {author} {\bibinfo {author} {\bibfnamefont {P.}~\bibnamefont
  {Hohenberg}}\ and\ \bibinfo {author} {\bibfnamefont {W.}~\bibnamefont
  {Kohn}},\ }\href@noop {} {\bibfield  {journal} {\bibinfo  {journal} {Phys.
  Rev.}\ }\textbf {\bibinfo {volume} {136}},\ \bibinfo {pages} {B864} (\bibinfo
  {year} {1964})}\BibitemShut {NoStop}%
\bibitem [{\citenamefont {Kohn}\ and\ \citenamefont {Sham}(1965)}]{Kohn1965}%
  \BibitemOpen
  \bibfield  {author} {\bibinfo {author} {\bibfnamefont {W.}~\bibnamefont
  {Kohn}}\ and\ \bibinfo {author} {\bibfnamefont {L.~J.}\ \bibnamefont
  {Sham}},\ }\href@noop {} {\bibfield  {journal} {\bibinfo  {journal} {Phys.
  Rev.}\ }\textbf {\bibinfo {volume} {140}},\ \bibinfo {pages} {A1133}
  (\bibinfo {year} {1965})}\BibitemShut {NoStop}%
\bibitem [{\citenamefont {Stefanucci}\ and\ \citenamefont
  {Almbladh}(2004)}]{Stefanucci2004}%
  \BibitemOpen
  \bibfield  {author} {\bibinfo {author} {\bibfnamefont {G.}~\bibnamefont
  {Stefanucci}}\ and\ \bibinfo {author} {\bibfnamefont {C.-O.}\ \bibnamefont
  {Almbladh}},\ }\href@noop {} {\bibfield  {journal} {\bibinfo  {journal}
  {Europhys. Lett.}\ }\textbf {\bibinfo {volume} {67}},\ \bibinfo {pages} {14}
  (\bibinfo {year} {2004})}\BibitemShut {NoStop}%
\bibitem [{\citenamefont {DiVentra}\ \emph {et~al.}(2000)\citenamefont
  {DiVentra}, \citenamefont {Pantelides},\ and\ \citenamefont
  {Lang}}]{DiVentra2000}%
  \BibitemOpen
  \bibfield  {author} {\bibinfo {author} {\bibfnamefont {M.}~\bibnamefont
  {DiVentra}}, \bibinfo {author} {\bibfnamefont {S.~T.}\ \bibnamefont
  {Pantelides}}, \ and\ \bibinfo {author} {\bibfnamefont {N.~D.}\ \bibnamefont
  {Lang}},\ }\href@noop {} {\bibfield  {journal} {\bibinfo  {journal} {Phys.
  Rev. Lett.}\ }\textbf {\bibinfo {volume} {84}},\ \bibinfo {pages} {979}
  (\bibinfo {year} {2000})}\BibitemShut {NoStop}%
\bibitem [{\citenamefont {Xue}\ \emph {et~al.}(2001)\citenamefont {Xue},
  \citenamefont {Datta},\ and\ \citenamefont {Ratner}}]{Xue2001}%
  \BibitemOpen
  \bibfield  {author} {\bibinfo {author} {\bibfnamefont {Y.}~\bibnamefont
  {Xue}}, \bibinfo {author} {\bibfnamefont {S.}~\bibnamefont {Datta}}, \ and\
  \bibinfo {author} {\bibfnamefont {M.~A.}\ \bibnamefont {Ratner}},\
  }\href@noop {} {\bibfield  {journal} {\bibinfo  {journal} {J. Chem. Phys.}\
  }\textbf {\bibinfo {volume} {115}},\ \bibinfo {pages} {4292} (\bibinfo {year}
  {2001})}\BibitemShut {NoStop}%
\bibitem [{\citenamefont {Xue}\ \emph {et~al.}(2002)\citenamefont {Xue},
  \citenamefont {Datta},\ and\ \citenamefont {Ratner}}]{Xue2002}%
  \BibitemOpen
  \bibfield  {author} {\bibinfo {author} {\bibfnamefont {Y.}~\bibnamefont
  {Xue}}, \bibinfo {author} {\bibfnamefont {S.}~\bibnamefont {Datta}}, \ and\
  \bibinfo {author} {\bibfnamefont {M.~A.}\ \bibnamefont {Ratner}},\
  }\href@noop {} {\bibfield  {journal} {\bibinfo  {journal} {Chem. Phys.}\
  }\textbf {\bibinfo {volume} {281}},\ \bibinfo {pages} {151} (\bibinfo {year}
  {2002})}\BibitemShut {NoStop}%
\bibitem [{\citenamefont {G\"{o}rling}\ and\ \citenamefont
  {Levy}(1995)}]{Gorling1995}%
  \BibitemOpen
  \bibfield  {author} {\bibinfo {author} {\bibfnamefont {A.}~\bibnamefont
  {G\"{o}rling}}\ and\ \bibinfo {author} {\bibfnamefont {M.}~\bibnamefont
  {Levy}},\ }\href@noop {} {\bibfield  {journal} {\bibinfo  {journal} {Phys.
  Rev. A}\ }\textbf {\bibinfo {volume} {52}},\ \bibinfo {pages} {4493}
  (\bibinfo {year} {1995})}\BibitemShut {NoStop}%
\bibitem [{\citenamefont {Evers}\ \emph {et~al.}(2004)\citenamefont {Evers},
  \citenamefont {Weigend},\ and\ \citenamefont {Koentopp}}]{Evers2004}%
  \BibitemOpen
  \bibfield  {author} {\bibinfo {author} {\bibfnamefont {F.}~\bibnamefont
  {Evers}}, \bibinfo {author} {\bibfnamefont {F.}~\bibnamefont {Weigend}}, \
  and\ \bibinfo {author} {\bibfnamefont {M.}~\bibnamefont {Koentopp}},\
  }\href@noop {} {\bibfield  {journal} {\bibinfo  {journal} {Phys. Rev. B}\
  }\textbf {\bibinfo {volume} {69}},\ \bibinfo {pages} {235411} (\bibinfo
  {year} {2004})}\BibitemShut {NoStop}%
\bibitem [{\citenamefont {Koentopp}\ \emph {et~al.}(2008)\citenamefont
  {Koentopp}, \citenamefont {Chang}, \citenamefont {Burke},\ and\ \citenamefont
  {Car}}]{Koentopp2008}%
  \BibitemOpen
  \bibfield  {author} {\bibinfo {author} {\bibfnamefont {M.}~\bibnamefont
  {Koentopp}}, \bibinfo {author} {\bibfnamefont {C.}~\bibnamefont {Chang}},
  \bibinfo {author} {\bibfnamefont {K.}~\bibnamefont {Burke}}, \ and\ \bibinfo
  {author} {\bibfnamefont {R.}~\bibnamefont {Car}},\ }\href@noop {} {\bibfield
  {journal} {\bibinfo  {journal} {J. Phys. Condens. Mat.}\ }\textbf {\bibinfo
  {volume} {20}},\ \bibinfo {pages} {083203} (\bibinfo {year}
  {2008})}\BibitemShut {NoStop}%
\bibitem [{\citenamefont {Yabana}\ and\ \citenamefont
  {Bertsch}(1996)}]{Yabana1996}%
  \BibitemOpen
  \bibfield  {author} {\bibinfo {author} {\bibfnamefont {K.}~\bibnamefont
  {Yabana}}\ and\ \bibinfo {author} {\bibfnamefont {G.~F.}\ \bibnamefont
  {Bertsch}},\ }\href@noop {} {\bibfield  {journal} {\bibinfo  {journal} {Phys.
  Rev. B}\ }\textbf {\bibinfo {volume} {54}},\ \bibinfo {pages} {4484}
  (\bibinfo {year} {1996})}\BibitemShut {NoStop}%
\bibitem [{\citenamefont {Li}\ \emph {et~al.}(2005)\citenamefont {Li},
  \citenamefont {Tully}, \citenamefont {Schlegel},\ and\ \citenamefont
  {Frisch}}]{Li2005}%
  \BibitemOpen
  \bibfield  {author} {\bibinfo {author} {\bibfnamefont {X.}~\bibnamefont
  {Li}}, \bibinfo {author} {\bibfnamefont {J.~C.}\ \bibnamefont {Tully}},
  \bibinfo {author} {\bibfnamefont {H.~B.}\ \bibnamefont {Schlegel}}, \ and\
  \bibinfo {author} {\bibfnamefont {M.~J.}\ \bibnamefont {Frisch}},\
  }\href@noop {} {\bibfield  {journal} {\bibinfo  {journal} {J. Chem. Phys.}\
  }\textbf {\bibinfo {volume} {123}},\ \bibinfo {pages} {084106} (\bibinfo
  {year} {2005})}\BibitemShut {NoStop}%
\bibitem [{\citenamefont {Qian}\ \emph {et~al.}(2006)\citenamefont {Qian},
  \citenamefont {Li}, \citenamefont {Lin},\ and\ \citenamefont
  {Yip}}]{Qian2006}%
  \BibitemOpen
  \bibfield  {author} {\bibinfo {author} {\bibfnamefont {X.}~\bibnamefont
  {Qian}}, \bibinfo {author} {\bibfnamefont {J.}~\bibnamefont {Li}}, \bibinfo
  {author} {\bibfnamefont {X.}~\bibnamefont {Lin}}, \ and\ \bibinfo {author}
  {\bibfnamefont {S.}~\bibnamefont {Yip}},\ }\href@noop {} {\bibfield
  {journal} {\bibinfo  {journal} {Phys. Rev. B}\ }\textbf {\bibinfo {volume}
  {73}},\ \bibinfo {pages} {035408} (\bibinfo {year} {2006})}\BibitemShut
  {NoStop}%
\bibitem [{\citenamefont {Varga}(2011)}]{Varga2011}%
  \BibitemOpen
  \bibfield  {author} {\bibinfo {author} {\bibfnamefont {K.}~\bibnamefont
  {Varga}},\ }\href@noop {} {\bibfield  {journal} {\bibinfo  {journal} {Phys.
  Rev. B}\ }\textbf {\bibinfo {volume} {83}},\ \bibinfo {pages} {195130}
  (\bibinfo {year} {2011})}\BibitemShut {NoStop}%
\bibitem [{\citenamefont {Neuhasuer}\ and\ \citenamefont
  {Baer}(1989)}]{Neuhasuer1989}%
  \BibitemOpen
  \bibfield  {author} {\bibinfo {author} {\bibfnamefont {D.}~\bibnamefont
  {Neuhasuer}}\ and\ \bibinfo {author} {\bibfnamefont {M.}~\bibnamefont
  {Baer}},\ }\href@noop {} {\bibfield  {journal} {\bibinfo  {journal} {J. Chem.
  Phys.}\ }\textbf {\bibinfo {volume} {90}},\ \bibinfo {pages} {4351} (\bibinfo
  {year} {1989})}\BibitemShut {NoStop}%
\bibitem [{\citenamefont {Vibok}\ and\ \citenamefont
  {Balint-Kurti}(1992{\natexlab{a}})}]{Vibok1992}%
  \BibitemOpen
  \bibfield  {author} {\bibinfo {author} {\bibfnamefont {A.}~\bibnamefont
  {Vibok}}\ and\ \bibinfo {author} {\bibfnamefont {G.~G.}\ \bibnamefont
  {Balint-Kurti}},\ }\href@noop {} {\bibfield  {journal} {\bibinfo  {journal}
  {J. Phys. Chem.}\ }\textbf {\bibinfo {volume} {96}},\ \bibinfo {pages} {8712}
  (\bibinfo {year} {1992}{\natexlab{a}})}\BibitemShut {NoStop}%
\bibitem [{\citenamefont {Vibok}\ and\ \citenamefont
  {Balint-Kurti}(1992{\natexlab{b}})}]{Vibok1992b}%
  \BibitemOpen
  \bibfield  {author} {\bibinfo {author} {\bibfnamefont {A.}~\bibnamefont
  {Vibok}}\ and\ \bibinfo {author} {\bibfnamefont {G.~G.}\ \bibnamefont
  {Balint-Kurti}},\ }\href@noop {} {\bibfield  {journal} {\bibinfo  {journal}
  {J. Chem. Phys.}\ }\textbf {\bibinfo {volume} {96}},\ \bibinfo {pages} {7615}
  (\bibinfo {year} {1992}{\natexlab{b}})}\BibitemShut {NoStop}%
\bibitem [{\citenamefont {Brouard}\ \emph {et~al.}(1994)\citenamefont
  {Brouard}, \citenamefont {Macias},\ and\ \citenamefont {Muga}}]{Brouard1994}%
  \BibitemOpen
  \bibfield  {author} {\bibinfo {author} {\bibfnamefont {S.}~\bibnamefont
  {Brouard}}, \bibinfo {author} {\bibfnamefont {D.}~\bibnamefont {Macias}}, \
  and\ \bibinfo {author} {\bibfnamefont {J.~G.}\ \bibnamefont {Muga}},\
  }\href@noop {} {\bibfield  {journal} {\bibinfo  {journal} {J. Phys. A: Math.
  Gen.}\ }\textbf {\bibinfo {volume} {27}},\ \bibinfo {pages} {L439} (\bibinfo
  {year} {1994})}\BibitemShut {NoStop}%
\bibitem [{\citenamefont {Riss}\ and\ \citenamefont {Meyer}(1996)}]{Riss1996}%
  \BibitemOpen
  \bibfield  {author} {\bibinfo {author} {\bibfnamefont {U.~V.}\ \bibnamefont
  {Riss}}\ and\ \bibinfo {author} {\bibfnamefont {H.-D.}\ \bibnamefont
  {Meyer}},\ }\href@noop {} {\bibfield  {journal} {\bibinfo  {journal} {J.
  Chem. Phys.}\ }\textbf {\bibinfo {volume} {105}},\ \bibinfo {pages} {1409}
  (\bibinfo {year} {1996})}\BibitemShut {NoStop}%
\bibitem [{\citenamefont {Ge}\ and\ \citenamefont {Zhang}(1997)}]{Ge1997}%
  \BibitemOpen
  \bibfield  {author} {\bibinfo {author} {\bibfnamefont {J.-Y.}\ \bibnamefont
  {Ge}}\ and\ \bibinfo {author} {\bibfnamefont {J.~Z.~H.}\ \bibnamefont
  {Zhang}},\ }\href@noop {} {\bibfield  {journal} {\bibinfo  {journal} {J.
  Chem. Phys.}\ }\textbf {\bibinfo {volume} {108}},\ \bibinfo {pages} {1429}
  (\bibinfo {year} {1997})}\BibitemShut {NoStop}%
\bibitem [{\citenamefont {Palao}\ \emph {et~al.}(1998)\citenamefont {Palao},
  \citenamefont {Muga},\ and\ \citenamefont {Sala}}]{Palao1998}%
  \BibitemOpen
  \bibfield  {author} {\bibinfo {author} {\bibfnamefont {J.~P.}\ \bibnamefont
  {Palao}}, \bibinfo {author} {\bibfnamefont {J.~G.}\ \bibnamefont {Muga}}, \
  and\ \bibinfo {author} {\bibfnamefont {R.}~\bibnamefont {Sala}},\ }\href@noop
  {} {\bibfield  {journal} {\bibinfo  {journal} {Phys. Rev. Lett.}\ }\textbf
  {\bibinfo {volume} {80}},\ \bibinfo {pages} {5469} (\bibinfo {year}
  {1998})}\BibitemShut {NoStop}%
\bibitem [{\citenamefont {Ferry}\ and\ \citenamefont
  {Barker}(1999)}]{Ferry1999}%
  \BibitemOpen
  \bibfield  {author} {\bibinfo {author} {\bibfnamefont {D.~K.}\ \bibnamefont
  {Ferry}}\ and\ \bibinfo {author} {\bibfnamefont {J.~R.}\ \bibnamefont
  {Barker}},\ }\href@noop {} {\bibfield  {journal} {\bibinfo  {journal} {Appl.
  Phys. Lett.}\ }\textbf {\bibinfo {volume} {74}},\ \bibinfo {pages} {582}
  (\bibinfo {year} {1999})}\BibitemShut {NoStop}%
\bibitem [{\citenamefont {Manolopoulos}(2002)}]{Manolopoulos2002}%
  \BibitemOpen
  \bibfield  {author} {\bibinfo {author} {\bibfnamefont {D.~E.}\ \bibnamefont
  {Manolopoulos}},\ }\href@noop {} {\bibfield  {journal} {\bibinfo  {journal}
  {J. Chem. Phys.}\ }\textbf {\bibinfo {volume} {117}},\ \bibinfo {pages}
  {9552} (\bibinfo {year} {2002})}\BibitemShut {NoStop}%
\bibitem [{\citenamefont {Poirier}\ and\ \citenamefont
  {Carrington}(2003{\natexlab{a}})}]{Poirier2003}%
  \BibitemOpen
  \bibfield  {author} {\bibinfo {author} {\bibfnamefont {B.}~\bibnamefont
  {Poirier}}\ and\ \bibinfo {author} {\bibfnamefont {T.}~\bibnamefont
  {Carrington}},\ }\href@noop {} {\bibfield  {journal} {\bibinfo  {journal} {J.
  Chem. Phys.}\ }\textbf {\bibinfo {volume} {118}},\ \bibinfo {pages} {17}
  (\bibinfo {year} {2003}{\natexlab{a}})}\BibitemShut {NoStop}%
\bibitem [{\citenamefont {Poirier}\ and\ \citenamefont
  {Carrington}(2003{\natexlab{b}})}]{Poirier2003b}%
  \BibitemOpen
  \bibfield  {author} {\bibinfo {author} {\bibfnamefont {B.}~\bibnamefont
  {Poirier}}\ and\ \bibinfo {author} {\bibfnamefont {T.}~\bibnamefont
  {Carrington}},\ }\href@noop {} {\bibfield  {journal} {\bibinfo  {journal} {J.
  Chem. Phys.}\ }\textbf {\bibinfo {volume} {119}},\ \bibinfo {pages} {77}
  (\bibinfo {year} {2003}{\natexlab{b}})}\BibitemShut {NoStop}%
\bibitem [{\citenamefont {Muga}\ \emph {et~al.}(2004)\citenamefont {Muga},
  \citenamefont {Palao}, \citenamefont {Navarro},\ and\ \citenamefont
  {Egusquiza}}]{Muga2004}%
  \BibitemOpen
  \bibfield  {author} {\bibinfo {author} {\bibfnamefont {J.~G.}\ \bibnamefont
  {Muga}}, \bibinfo {author} {\bibfnamefont {J.~P.}\ \bibnamefont {Palao}},
  \bibinfo {author} {\bibfnamefont {B.}~\bibnamefont {Navarro}}, \ and\
  \bibinfo {author} {\bibfnamefont {I.~L.}\ \bibnamefont {Egusquiza}},\
  }\href@noop {} {\bibfield  {journal} {\bibinfo  {journal} {Phys. Rep.}\
  }\textbf {\bibinfo {volume} {395}},\ \bibinfo {pages} {357} (\bibinfo {year}
  {2004})}\BibitemShut {NoStop}%
\bibitem [{\citenamefont {Kosloff}\ and\ \citenamefont
  {Kosloff}(1986)}]{Kosloff1986}%
  \BibitemOpen
  \bibfield  {author} {\bibinfo {author} {\bibfnamefont {R.}~\bibnamefont
  {Kosloff}}\ and\ \bibinfo {author} {\bibfnamefont {D.}~\bibnamefont
  {Kosloff}},\ }\href@noop {} {\bibfield  {journal} {\bibinfo  {journal} {J.
  Comput. Phys.}\ }\textbf {\bibinfo {volume} {63}},\ \bibinfo {pages} {363}
  (\bibinfo {year} {1986})}\BibitemShut {NoStop}%
\bibitem [{\citenamefont {Driscoll}\ and\ \citenamefont
  {Varga}(2008)}]{Driscoll2008}%
  \BibitemOpen
  \bibfield  {author} {\bibinfo {author} {\bibfnamefont {J.~A.}\ \bibnamefont
  {Driscoll}}\ and\ \bibinfo {author} {\bibfnamefont {K.}~\bibnamefont
  {Varga}},\ }\href@noop {} {\bibfield  {journal} {\bibinfo  {journal} {Phys.
  Rev. B}\ }\textbf {\bibinfo {volume} {78}},\ \bibinfo {pages} {245118}
  (\bibinfo {year} {2008})}\BibitemShut {NoStop}%
\bibitem [{\citenamefont {Varga}(2009)}]{Varga2009}%
  \BibitemOpen
  \bibfield  {author} {\bibinfo {author} {\bibfnamefont {K.}~\bibnamefont
  {Varga}},\ }\href@noop {} {\bibfield  {journal} {\bibinfo  {journal} {Phys.
  Status Solidi B}\ }\textbf {\bibinfo {volume} {246}},\ \bibinfo {pages}
  {1407} (\bibinfo {year} {2009})}\BibitemShut {NoStop}%
\bibitem [{\citenamefont {Feldman}\ \emph {et~al.}(2014)\citenamefont
  {Feldman}, \citenamefont {Siedeman}, \citenamefont {Hod},\ and\ \citenamefont
  {Kronik}}]{Feldman2014}%
  \BibitemOpen
  \bibfield  {author} {\bibinfo {author} {\bibfnamefont {B.}~\bibnamefont
  {Feldman}}, \bibinfo {author} {\bibfnamefont {T.}~\bibnamefont {Siedeman}},
  \bibinfo {author} {\bibfnamefont {O.}~\bibnamefont {Hod}}, \ and\ \bibinfo
  {author} {\bibfnamefont {L.}~\bibnamefont {Kronik}},\ }\href@noop {}
  {\bibfield  {journal} {\bibinfo  {journal} {Phys. Rev. B}\ }\textbf {\bibinfo
  {volume} {90}},\ \bibinfo {pages} {035445} (\bibinfo {year}
  {2014})}\BibitemShut {NoStop}%
\bibitem [{\citenamefont {Wibking}\ and\ \citenamefont
  {Varga}(2012)}]{Wibking2012}%
  \BibitemOpen
  \bibfield  {author} {\bibinfo {author} {\bibfnamefont {B.~D.}\ \bibnamefont
  {Wibking}}\ and\ \bibinfo {author} {\bibfnamefont {K.}~\bibnamefont
  {Varga}},\ }\href@noop {} {\bibfield  {journal} {\bibinfo  {journal} {Phys.
  Lett. A}\ }\textbf {\bibinfo {volume} {376}},\ \bibinfo {pages} {365}
  (\bibinfo {year} {2012})}\BibitemShut {NoStop}%
\bibitem [{\citenamefont {Varga}\ and\ \citenamefont
  {Pantelides}(2007)}]{Varga2007}%
  \BibitemOpen
  \bibfield  {author} {\bibinfo {author} {\bibfnamefont {K.}~\bibnamefont
  {Varga}}\ and\ \bibinfo {author} {\bibfnamefont {S.~T.}\ \bibnamefont
  {Pantelides}},\ }\href@noop {} {\bibfield  {journal} {\bibinfo  {journal}
  {Phys. Rev. Lett.}\ }\textbf {\bibinfo {volume} {98}},\ \bibinfo {pages}
  {076804} (\bibinfo {year} {2007})}\BibitemShut {NoStop}%
\bibitem [{\citenamefont {Wahlstrand}\ \emph {et~al.}(2014)\citenamefont
  {Wahlstrand}, \citenamefont {Yakimenko},\ and\ \citenamefont
  {Berggren}}]{Wahlstrand2014}%
  \BibitemOpen
  \bibfield  {author} {\bibinfo {author} {\bibfnamefont {B.}~\bibnamefont
  {Wahlstrand}}, \bibinfo {author} {\bibfnamefont {I.~I.}\ \bibnamefont
  {Yakimenko}}, \ and\ \bibinfo {author} {\bibfnamefont {K.-F.}\ \bibnamefont
  {Berggren}},\ }\href@noop {} {\bibfield  {journal} {\bibinfo  {journal}
  {Phys. Rev. E}\ }\textbf {\bibinfo {volume} {89}},\ \bibinfo {pages} {062910}
  (\bibinfo {year} {2014})}\BibitemShut {NoStop}%
\bibitem [{\citenamefont {Reddy}\ \emph {et~al.}(2012)\citenamefont {Reddy},
  \citenamefont {Jadaun}, \citenamefont {Valsaraj}, \citenamefont {Register},\
  and\ \citenamefont {Banerjee}}]{Reddy2012}%
  \BibitemOpen
  \bibfield  {author} {\bibinfo {author} {\bibfnamefont {D.}~\bibnamefont
  {Reddy}}, \bibinfo {author} {\bibfnamefont {P.}~\bibnamefont {Jadaun}},
  \bibinfo {author} {\bibfnamefont {A.}~\bibnamefont {Valsaraj}}, \bibinfo
  {author} {\bibfnamefont {L.~F.}\ \bibnamefont {Register}}, \ and\ \bibinfo
  {author} {\bibfnamefont {S.~K.}\ \bibnamefont {Banerjee}},\ }in\ \href@noop
  {} {\emph {\bibinfo {booktitle} {Proceedings of the International Conference
  on Simulation of Semiconductor Processes and Devices}}}\ (\bibinfo {year}
  {2012})\ pp.\ \bibinfo {pages} {51--54}\BibitemShut {NoStop}%
\bibitem [{\citenamefont {Umar}\ \emph {et~al.}(1985)\citenamefont {Umar},
  \citenamefont {Strayer}, \citenamefont {Cusson}, \citenamefont {Reinhard},\
  and\ \citenamefont {Bromley}}]{Umar1985}%
  \BibitemOpen
  \bibfield  {author} {\bibinfo {author} {\bibfnamefont {A.~S.}\ \bibnamefont
  {Umar}}, \bibinfo {author} {\bibfnamefont {M.~R.}\ \bibnamefont {Strayer}},
  \bibinfo {author} {\bibfnamefont {R.~Y.}\ \bibnamefont {Cusson}}, \bibinfo
  {author} {\bibfnamefont {P.-G.}\ \bibnamefont {Reinhard}}, \ and\ \bibinfo
  {author} {\bibfnamefont {D.~A.}\ \bibnamefont {Bromley}},\ }\href@noop {}
  {\bibfield  {journal} {\bibinfo  {journal} {Phys. Rev. C}\ }\textbf {\bibinfo
  {volume} {32}},\ \bibinfo {pages} {172} (\bibinfo {year} {1985})}\BibitemShut
  {NoStop}%
\bibitem [{\citenamefont {Wu}\ and\ \citenamefont {Voorhis}(2005)}]{Wu2005}%
  \BibitemOpen
  \bibfield  {author} {\bibinfo {author} {\bibfnamefont {Q.}~\bibnamefont
  {Wu}}\ and\ \bibinfo {author} {\bibfnamefont {T.~V.}\ \bibnamefont
  {Voorhis}},\ }\href@noop {} {\bibfield  {journal} {\bibinfo  {journal} {Phys.
  Rev. A}\ }\textbf {\bibinfo {volume} {72}},\ \bibinfo {pages} {024502}
  (\bibinfo {year} {2005})}\BibitemShut {NoStop}%
\bibitem [{\citenamefont {Oberhofer}\ and\ \citenamefont
  {Blumberger}(2009)}]{Oberhofer2009}%
  \BibitemOpen
  \bibfield  {author} {\bibinfo {author} {\bibfnamefont {H.}~\bibnamefont
  {Oberhofer}}\ and\ \bibinfo {author} {\bibfnamefont {J.}~\bibnamefont
  {Blumberger}},\ }\href@noop {} {\bibfield  {journal} {\bibinfo  {journal} {J.
  Chem. Phys.}\ }\textbf {\bibinfo {volume} {131}},\ \bibinfo {pages} {064101}
  (\bibinfo {year} {2009})}\BibitemShut {NoStop}%
\bibitem [{\citenamefont {Elenewski}\ and\ \citenamefont
  {Chen}(2014)}]{Elenewski2014}%
  \BibitemOpen
  \bibfield  {author} {\bibinfo {author} {\bibfnamefont {J.~E.}\ \bibnamefont
  {Elenewski}}\ and\ \bibinfo {author} {\bibfnamefont {H.}~\bibnamefont
  {Chen}},\ }\href@noop {} {\bibfield  {journal} {\bibinfo  {journal} {Phys.
  Rev. B}\ }\textbf {\bibinfo {volume} {90}},\ \bibinfo {pages} {085104}
  (\bibinfo {year} {2014})}\BibitemShut {NoStop}%
\bibitem [{\citenamefont {Imry}\ and\ \citenamefont
  {Landauer}(1999)}]{Imry1999}%
  \BibitemOpen
  \bibfield  {author} {\bibinfo {author} {\bibfnamefont {Y.}~\bibnamefont
  {Imry}}\ and\ \bibinfo {author} {\bibfnamefont {R.}~\bibnamefont
  {Landauer}},\ }\href@noop {} {\bibfield  {journal} {\bibinfo  {journal} {Rev.
  Mod. Phys.}\ }\textbf {\bibinfo {volume} {71}},\ \bibinfo {pages} {S306}
  (\bibinfo {year} {1999})}\BibitemShut {NoStop}%
\bibitem [{\citenamefont {Marques}\ \emph {et~al.}(2003)\citenamefont
  {Marques}, \citenamefont {Castro}, \citenamefont {Bertsch},\ and\
  \citenamefont {Rubio}}]{Marques2003}%
  \BibitemOpen
  \bibfield  {author} {\bibinfo {author} {\bibfnamefont {M.~A.~L.}\
  \bibnamefont {Marques}}, \bibinfo {author} {\bibfnamefont {A.}~\bibnamefont
  {Castro}}, \bibinfo {author} {\bibfnamefont {G.~F.}\ \bibnamefont {Bertsch}},
  \ and\ \bibinfo {author} {\bibfnamefont {A.}~\bibnamefont {Rubio}},\
  }\href@noop {} {\bibfield  {journal} {\bibinfo  {journal} {Comput. Phys.
  Commun.}\ }\textbf {\bibinfo {volume} {151}},\ \bibinfo {pages} {60}
  (\bibinfo {year} {2003})}\BibitemShut {NoStop}%
\bibitem [{\citenamefont {Castro}\ \emph {et~al.}(2004)\citenamefont {Castro},
  \citenamefont {Marques},\ and\ \citenamefont {Rubio}}]{Castro2004}%
  \BibitemOpen
  \bibfield  {author} {\bibinfo {author} {\bibfnamefont {A.}~\bibnamefont
  {Castro}}, \bibinfo {author} {\bibfnamefont {M.~A.~L.}\ \bibnamefont
  {Marques}}, \ and\ \bibinfo {author} {\bibfnamefont {A.}~\bibnamefont
  {Rubio}},\ }\href@noop {} {\bibfield  {journal} {\bibinfo  {journal} {J.
  Chem. Phys.}\ }\textbf {\bibinfo {volume} {121}},\ \bibinfo {pages} {3425}
  (\bibinfo {year} {2004})}\BibitemShut {NoStop}%
\bibitem [{\citenamefont {Castro}\ \emph {et~al.}(2006)\citenamefont {Castro},
  \citenamefont {Appel}, \citenamefont {Oliveira}, \citenamefont {Rozzi},
  \citenamefont {Andrade}, \citenamefont {Lorenzen}, \citenamefont {Marques},
  \citenamefont {Gross},\ and\ \citenamefont {Rubio}}]{Castro2006}%
  \BibitemOpen
  \bibfield  {author} {\bibinfo {author} {\bibfnamefont {A.}~\bibnamefont
  {Castro}}, \bibinfo {author} {\bibfnamefont {H.}~\bibnamefont {Appel}},
  \bibinfo {author} {\bibfnamefont {M.}~\bibnamefont {Oliveira}}, \bibinfo
  {author} {\bibfnamefont {C.~A.}\ \bibnamefont {Rozzi}}, \bibinfo {author}
  {\bibfnamefont {X.}~\bibnamefont {Andrade}}, \bibinfo {author} {\bibfnamefont
  {F.}~\bibnamefont {Lorenzen}}, \bibinfo {author} {\bibfnamefont {M.~A.~L.}\
  \bibnamefont {Marques}}, \bibinfo {author} {\bibfnamefont {E.~K.~U.}\
  \bibnamefont {Gross}}, \ and\ \bibinfo {author} {\bibfnamefont
  {A.}~\bibnamefont {Rubio}},\ }\href@noop {} {\bibfield  {journal} {\bibinfo
  {journal} {Phys. Stat. Sol. B}\ }\textbf {\bibinfo {volume} {243}},\ \bibinfo
  {pages} {2465} (\bibinfo {year} {2006})}\BibitemShut {NoStop}%
\bibitem [{\citenamefont {Andrade}\ \emph {et~al.}(2012)\citenamefont
  {Andrade}, \citenamefont {Alberdi-Rodriguez}, \citenamefont {Strubbe},
  \citenamefont {Oliveira}, \citenamefont {Nogueira}, \citenamefont {Castro},
  \citenamefont {Muguerza}, \citenamefont {Arruabarrena}, \citenamefont
  {Louie}, \citenamefont {Aspuru-Guzik}, \citenamefont {Rubio},\ and\
  \citenamefont {Marques}}]{Andrade2012}%
  \BibitemOpen
  \bibfield  {author} {\bibinfo {author} {\bibfnamefont {X.}~\bibnamefont
  {Andrade}}, \bibinfo {author} {\bibfnamefont {J.}~\bibnamefont
  {Alberdi-Rodriguez}}, \bibinfo {author} {\bibfnamefont {D.~A.}\ \bibnamefont
  {Strubbe}}, \bibinfo {author} {\bibfnamefont {M.~J.~T.}\ \bibnamefont
  {Oliveira}}, \bibinfo {author} {\bibfnamefont {F.}~\bibnamefont {Nogueira}},
  \bibinfo {author} {\bibfnamefont {A.}~\bibnamefont {Castro}}, \bibinfo
  {author} {\bibfnamefont {J.}~\bibnamefont {Muguerza}}, \bibinfo {author}
  {\bibfnamefont {A.}~\bibnamefont {Arruabarrena}}, \bibinfo {author}
  {\bibfnamefont {S.~G.}\ \bibnamefont {Louie}}, \bibinfo {author}
  {\bibfnamefont {A.}~\bibnamefont {Aspuru-Guzik}}, \bibinfo {author}
  {\bibfnamefont {A.}~\bibnamefont {Rubio}}, \ and\ \bibinfo {author}
  {\bibfnamefont {M.~A.~L.}\ \bibnamefont {Marques}},\ }\href@noop {}
  {\bibfield  {journal} {\bibinfo  {journal} {J. Phys.: Condens. Matt.}\
  }\textbf {\bibinfo {volume} {24}},\ \bibinfo {pages} {233202} (\bibinfo
  {year} {2012})}\BibitemShut {NoStop}%
\end{thebibliography}
\end{document}